\documentclass{article}
\usepackage[utf8]{inputenc}
\usepackage{charter} 
\usepackage{geometry}
\usepackage{authblk}
\usepackage{siunitx}
\usepackage{amsfonts}
\usepackage{hyperref}
\usepackage{graphicx, caption}
\usepackage{xcolor}
\usepackage[style=nature]{biblatex}
\usepackage{adjustbox}
\usepackage{tabularx}
\usepackage{mathtools}
\usepackage[version=3]{mhchem} 
\usepackage{amssymb}
\usepackage{braket}
\usepackage{esvect}
\usepackage{listings}
\usepackage{mathptmx}
\usepackage{amsmath}

\DeclareUnicodeCharacter{0300}{*************************************}

\graphicspath{ {./figures/} }

\title{Quantifying the Atomistic Free-volume Morphology of Materials with Graph Theory}

\author[1]{James Chapman \thanks{Corresponding Author, chapman37@llnl.gov}}
\author[1,2]{Nir Goldman}

\affil[1]{Materials Science Division, Lawrence Livermore National Laboratory, Livermore, CA, USA}
\affil[2]{Department of Chemical Engineering, University of California, Davis, California 95616, United States}

\begin{document}

\maketitle

\begin{abstract}
    We introduce a new computational methodology for the identification and characterization of free volume within/around atomistic configurations. This scheme employs a three-stage workflow, by which spheres are iteratively grown inside of voxels, and ultimately converted to planar graphs, which are then characterized via a graph-based order parameter. Our approach is computationally efficient, physically intuitive, and universally transferable to any material system. Validation of our methodology is performed on several sets of materials problems: (1) classification of unique free volumes in various crystal phases, (2) autonomous detection and classification of complex surface defects during epitaxial growth simulations, (3) characterization of free volume defects in metals/alloys, and (4) quantification of the spatio-temporal behavior of nano-scale free volume morphologies as a function of both temperature and free-volume size. Our method accurately identifies and characterizes unique free volumes over a multitude of systems and length scales, indicating its potential for future use in understanding the relationship between free volume morphology and material properties under both static and dynamic conditions.

\end{abstract}
\section{Introduction}

Characterizing the morphology of free volume in materials at the atomic level is critical to our understanding of the material's underlying mechanical and transport properties \cite{https://doi.org/10.1002/app.40376,doi:10.1021/ma0213448,MCGONIGLE20012413,HEDENQVIST19962887,doi:10.1021/ie9007503,https://doi.org/10.1002/pi.2289}. Macroscale phenomena such as fracture, phase transformations, and diffusivity are coupled to the structure of free volume within a material \cite{LAUNEY2008500,doi:10.1063/1.3447751,KUSY2006982,WANG2019151924}. For example, problems such as void swelling in crystalline materials due to radiation damage involves Frenkel pair formation and ultimately void coalescence within the bulk material, potentially leading to catastrophic failure \cite{WOO19967,doi:10.1080/01418610208240021}. Mass transport within polymeric systems is largely dictated by the local void structure, which can vary significantly over sample size and with enhanced thermodynamic conditions\cite{polymer_transport2016}. Porous organic materials such as covalent organic frameworks (COFs) have potential for a number of application areas, including energy storage, catalysis, and others\cite{porous_organic2016}. Consequently, a quantitative mapping technique that can uniquely identify free-volume structures within a material is vital for reliable elucidation of experimental studies and can potentially guide future synthesis by allowing for pore structure to be used as a design tool. 

However, this characterization is often non-trivial, especially for disordered systems in which the lack of atomic symmetry often makes determining free volume topology challenging. Throughout the past two decades, many techniques have been employed to directly characterize these morphologies such as Voronoi Site Detection (VSD) \cite{Stukowski_2012}, Landmark Analysis (LAND) \cite{PhysRevMaterials.3.055404}, Kaundinya et al's voxelization scheme (VOX) \cite{PhysRevMaterials.5.063802}, Minkowski Tensors (MT) \cite{https://doi.org/10.1002/adma.201100562}, Chan et al's machine learning (ML) algorithm for determining 3D microstructures (ML3D) \cite{Chan2020}, and Byska et al's Voronoi-based region of interest scheme (VROI) \cite{10.1145/2788539.2788548}. Methods that rely purely on voxelization of either atoms or their surrounding local volume, such as ML3D, and VOX, often lack information regarding voxel-atom connectivity (e.g., correlations over local spatial scales), making it challenging to uniquely characterize free-volumes and their effect on material properties.  In contrast to these methods, geometric manipulations such as Voronoi tessellations also form the foundation of many methodologies, such as VSD, VROI, MT, and LAND. However, Voronoi tessellations often require a reference system in order to be interpretable, potentially making it difficult to quantify previously uncharacterized free volume topologies for amorphous systems or phase changes in particular. Atomistic symmetry functions, such as the Smooth Overlap of Atomic Positions (SOAP) \cite{PhysRevB.87.184115} and the Behler-Parrinello fingerprints \cite{doi:10.1063/1.3553717} lack long-range connectivity information, resulting in their inability to capture the shape and density of features such as voids, channels, clusters, etc. \cite{chapman2021sgop}.

In this work, we overcome these challenges through the development of a physically intuitive and computationally efficient algorithm, referred to as the Parallelized Atomistic Nanoscale Defect Analyzer (PANDA). Our approach leverages voxelization, due to its highly parallelized nature, and places infinitesimally small spheres at the center of each voxel. These spheres are then iteratively grown until a set of user-defined rules has been satisfied. A handful of physically informed parameters are used to drive the algorithm, though all can be derived from the local geometric order present in the system, such as the radial distribution function (RDF). The spheres are then mapped onto a planar graph, where free volume identification and characterization are performed. The threefold workflow of voxels to spheres to planar graph ensures that the final free volume topology is encoded in a physically informed way. 

\begin{figure*}
        \centering
    	\includegraphics[trim={0 0cm 0 0cm},width=1.0\textwidth]{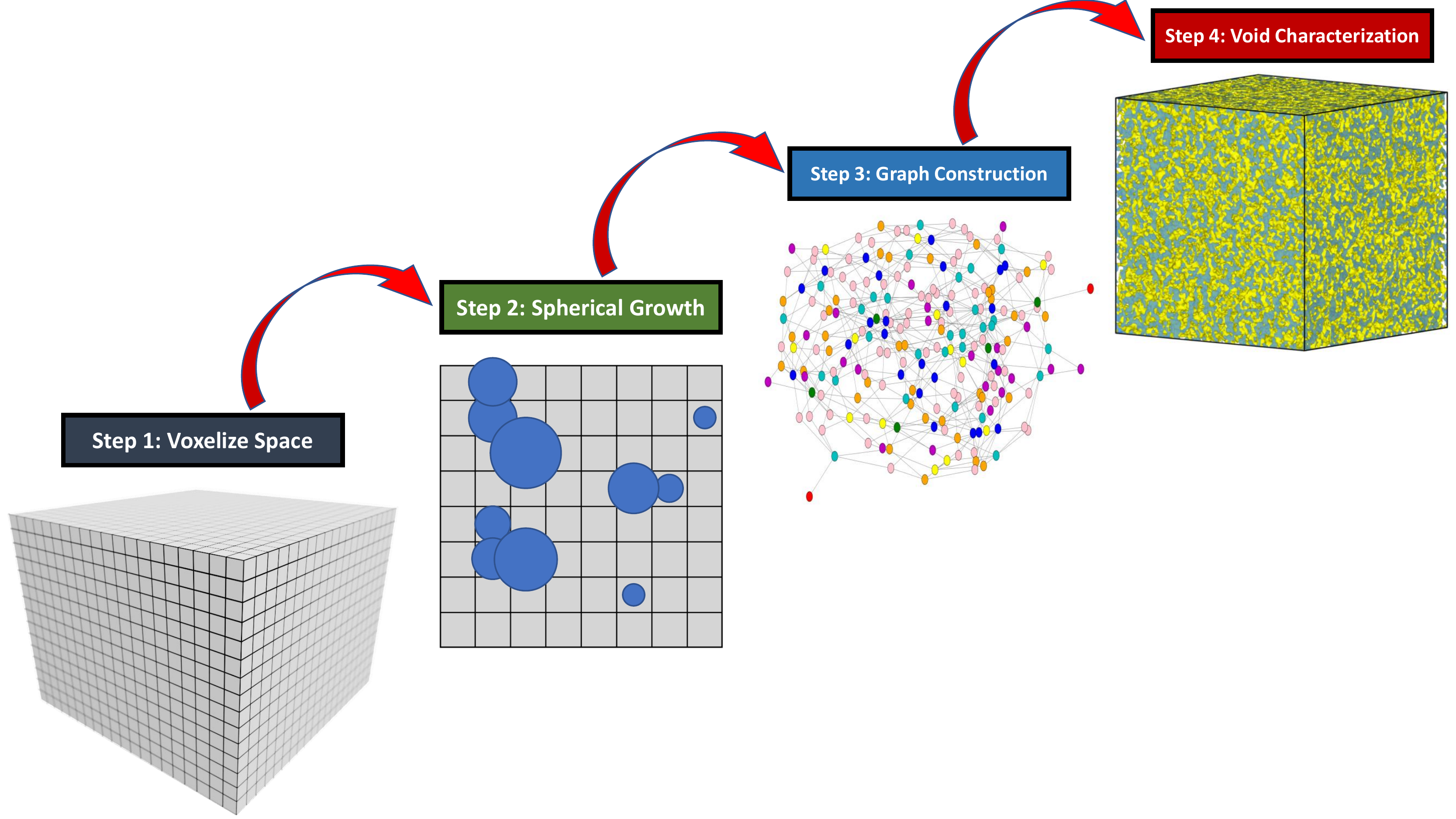}
    	\caption{Visual depiction of the algorithmic workflow of PANDA. (left) The input space is voxelized and spheres of infinitesimal radius are placed at voxel centers. Spheres are then grown iteratively, until user-defined rulesets have been satisfied (shown here in 2D, with blue circles representing their 3D counterparts). Planar graphs are constructed from the overlaps between spheres, with vertex information derived from a sphere's local connectivity. Unsupervised clustering and characterization is then performed on the planar graph to obtain the free-volume morphology of the input system.}
    	\label{fig:theory}
\end{figure*}

The remainder of the manuscript is as follows. First, we provide the reader with a detailed understanding of the theoretical framework that drives the PANDA algorithm. Next, we demonstrate how the free volume of various crystal phases of aluminum can be obtained by PANDA. Complex surface features are then identified and characterized after dynamic epitixial growth simulations on the Al (110) surface. Various types of free volume defects are then analyzed, such as mono, di, and trivacancies, as well the differences between vacancies created by removing unique chemical species in multicomponent systems. We highlight PANDA's ability to autonomously track vacancy diffusion in aluminum during molecular dynamics (MD) simulations. We then showcase how large-scale characterization of free volume morphology can be observed by determining the various types of defects in Al. Finally, we provide detailed insight into the spatio-temporal behavior of different free volume morphologies in Al systems containing more than a million atoms through nanosecond MD simulations, and connect these findings with experimental observations.

\section{Computational Details}

\subsection{PANDA Algorithm}

The PANDA workflow is broken into several pieces: (1) the total volume of the input space is voxelized into smaller subspaces, (2) spheres are placed at the center of each voxel and grown iteratively under a given ruleset is satisfied, and (3) planar graphs are constructed from the overlaps between spheres. Figure \ref{fig:theory} provides a visual depiction of this workflow. Stage 1 takes the input space and voxelizes it, creating a computationally cheap and efficient way of discretizing the configuration. Here we define voxel neighbors as adjacent voxels (nearest-neighbors), allowing for an even further discretization of the input space. As the voxel neighbors are predefined by nature, one does not need to consider non-local information during graph construction (stage 3), improving the scaling of the algorithm. 

Stage 2 places spheres of arbitrarily small radius at the center of each voxel. These spheres are then allowed to grow, iteratively, until all desired rules have been satisfied. The rulesets that govern sphere growth are, by nature, user-defined, providing one with the ability to tailor the growth behavior to capture a desired aspect of free-volume morphology. In this work, the growth of our spheres employs a simple set of rules which govern when to turn off the growth of a specific sphere: (a) $r_{s} \ge 2l_{v}$, where $r_{s}$ is the radius of a given sphere and $l_{v}$ is the length of a voxel, (b) if any point on the sphere falls within a user-defined cutoff radius, $R_{c}^{0}$ of an atom, which is defined uniquely for each chemical species present in the system (e.g., the van der Waals radius), and (c) if there are no atoms within a user-defined cutoff radius of a voxel center, $R_{c}^{1}$ (which in this work was set to $2R_{c}^{0}$). Step (c) is checked once during the setup of the algorithm, as the atomic positions are held fixed during the calculation, while steps (a) and (b) are checked during each growth step.

After all of the growth has been switched off, a planar graph is constructed from the resulting spherical overlaps. As each voxel only has a predefined number of neighbors, sphere overlaps are checked only within this voxel neighbor list. If any point on a given sphere falls within a neighboring sphere the two voxels are joined together by an edge on the planar graph, while the volume of the sphere is recorded at each node in the graph. A Voxel whose sphere does not overlap with any of its neighbors is not counted in the graph. The resulting planar graph is then used to perform unsupervised clustering of the free-volume in the system by determining the subgraphs contained with the graph. Each subgraph represents a unique free-volume polyhedra, whose total volume is simply the sum of the node values within the subgraph. 

Overall, we find that PANDA yields strong computational efficiency for analysis of these types of simulation supercells. For the largest systems studied in this work, measured at 25 nanometers cubed, and contained over 1 million atoms, the PANDA free volume map was generated in less than thirty seconds using 36 threads on an Intel Xeon E5-2695 v4 processor. This time can potentially be reduced further by exploring a more efficient voxel-to-atom neighbor list generation scheme, as well as opting for a distributed memory approach rather than a shared memory one.

\subsection{Graph Characterization}

In this work we use two methods for characterizing graph morphology. The first is simply the total volume of a given subgraph, defined in the previous section. However, in many cases, the volume of the subgraph may not be sufficient to uniquely identify subtle differences between free-volumes, due to the inter-connectivity between voxels (e.g., linear vs. three-dimensional morphologies). Therefore, we employ a  graph-based adjacency matrix to represent the spherical overlaps present in the system. At the node-level of the graph, the degree of each node is simply the number of voxel-voxel sphere overlaps. SGOP-V, which is a modification of a previous graph-based order parameter \cite{chapman2021sgop}, is then defined as:

\begin{equation}
    \theta_{G} = \frac{1}{V} \sum_{s}^{S} \sum_{k}^{\omega_{s}}\frac{4}{3}\pi r_{k}^{3}\left( \sum_{m}^{D_{s}}P(d_{m})\log_{b}P(d_{m}) + d_{m}P(d_{m})  \right)^{3}
\end{equation}

Each underlying graph network exists as a set of subgraphs, $S$, with $s$ indexing a particular subgraph. $\omega_{i}$ is the set of all nodes (voxels) contained within a given subgraph $s$. $r_{k}$ is the radius of the sphere located at voxel $k$. $D_{s}$ is the set of unique node degrees in a subgraph, with $P_{d_{m}}$ being the probability of a given degree, $d_{m}$, occurring in the subgraph. $V$ is the total volume of the input space, which acts as a normalization term, allowing for the direct comparison of free-volume morphologies regardless of the size of the structure. SGOP-V is then used to distinguish unique classes of free-volume topologies. Further information regarding the theory behind this methodology can be found in our previous work \cite{chapman2021sgop}.

\subsection{Event Tracking}

Event tracking is defined as the ability to track positional changes of free-volumes during a dynamic trajectory. A simple way to visualize this would be to imagine a vacancy hopping to an adjacent site in a bulk material. Rather than tracking the hopping of an atom to a vacancy site, we can define the reverse, in which we track the movement of the vacancy site itself. Such a scheme becomes extremely useful when tracking the motion of large-scale voids under dynamic conditions, though in this work we limit ourselves to tracking the motion of a single vacancy in an otherwise pristine bulk system, as a proof of concept. As the graph nodes described in the previous sections represent voxels in a three-dimensional space, we can assign real-space positions to each node in the graph. By doing this, one can now track the COM of a subgraph, defined as:

\begin{equation}
	R_{s} = \frac{1}{M_{s}} \sum_{k}^{\omega_{s}}d_{k}r_{k}
\end{equation}

Here $R_{s}$ is defined as the COM of a particular subgraph. $M_{s}$ represents the fictitious mass of the subgraph, defined as $\sum_{i}^{D_{s}}d_{i}$, where $D_{s}$ is the set of all degrees in subgraph $s$, and $d_{i}$ represents each individual degree. Similar to our previous discussion, $\omega_{s}$ is the set of all nodes in the subgraph $s$, and $d_{k}$ and $r_{k}$ are the degree of each node, and the position of the node (voxel), respectively. Use of the degree to calculate the fictitious mass of the subgraph allows for the COM to be determined from the connectivity present in the subgraph as well as its shape.

We then define a change of position of the subgraph as the change in the subgraph's COM, $\delta R_{s}$. If $\delta R_{s}$ is greater than a pre-defined tolerance, $\delta_{tol}$, then we say that the subgraph has moved to a new position. We define this tolerance in this work to be $\delta_{tol} \ge Cl_{v}$. We note that the scaling factor $C$ is relative to $l_{v}$, with a finer voxel mesh requiring a larger scaling factor. For this work, we find a scaling factor of 3.67 (which equates to a $\delta_{tol}$ of 3\AA) is appropriate. Over the course of several dynamic trajectories calculated at various temperatures, we track the time between subgraph hops to find a relationship between the hop rates and the temperature. This relationship allows us the use an Arrhenius fit to extract the activation energy and prefactor required for the vacancy to hop between adjacent sites. Our computed activation energy was then compared to the zero-temperature activation energy, calculated via the nudged elastic band (NEB) algorithm. All calculations performed in this work were done on elemental aluminum using the LAMMPS software suite\cite{PLIMPTON19951}.

\subsection{Graph Contouring}

Graph contouring is the process of reducing a graph based on its real-space positions to encode information about the length, width, and height of features captured by the iterative sphere growth. We can leverage step (c) of our sphere growth ruleset (i.e., to turn off growth for spheres that are far away from an atom) to readily identify surface features, in a way similar to pressing the surface into a mold to understand its shape. We define a $z$-plane as the surface level, in which any free-volume below this point is ignored for practical reasons. Subgraphs are then determined on the remaining nodes, with each node encoding the real-space position of its host voxel.

The subgraphs are then projected back into real-space and the space is binned along the $x$ and $y$ directions. A contouring algorithm is then used to map the boundaries of each binned-subgraph, with the length and width of each contour defined as the largest distance between points along the contour line, in the specified direction. This distance, however, is defined in the binned region, and is therefore projected back into real-space. These final two values represent the length and width of surface features present in the system. One can control the fidelity of these values by adjusting the number of bins along each direction in the real-space binning of the subgraphs.
\section{Results}

\subsection{Atomic Packing Factor Characterization}

\begin{figure*}
        \centering
    	\includegraphics[trim={0 0cm 0 0cm},width=0.8\textwidth]{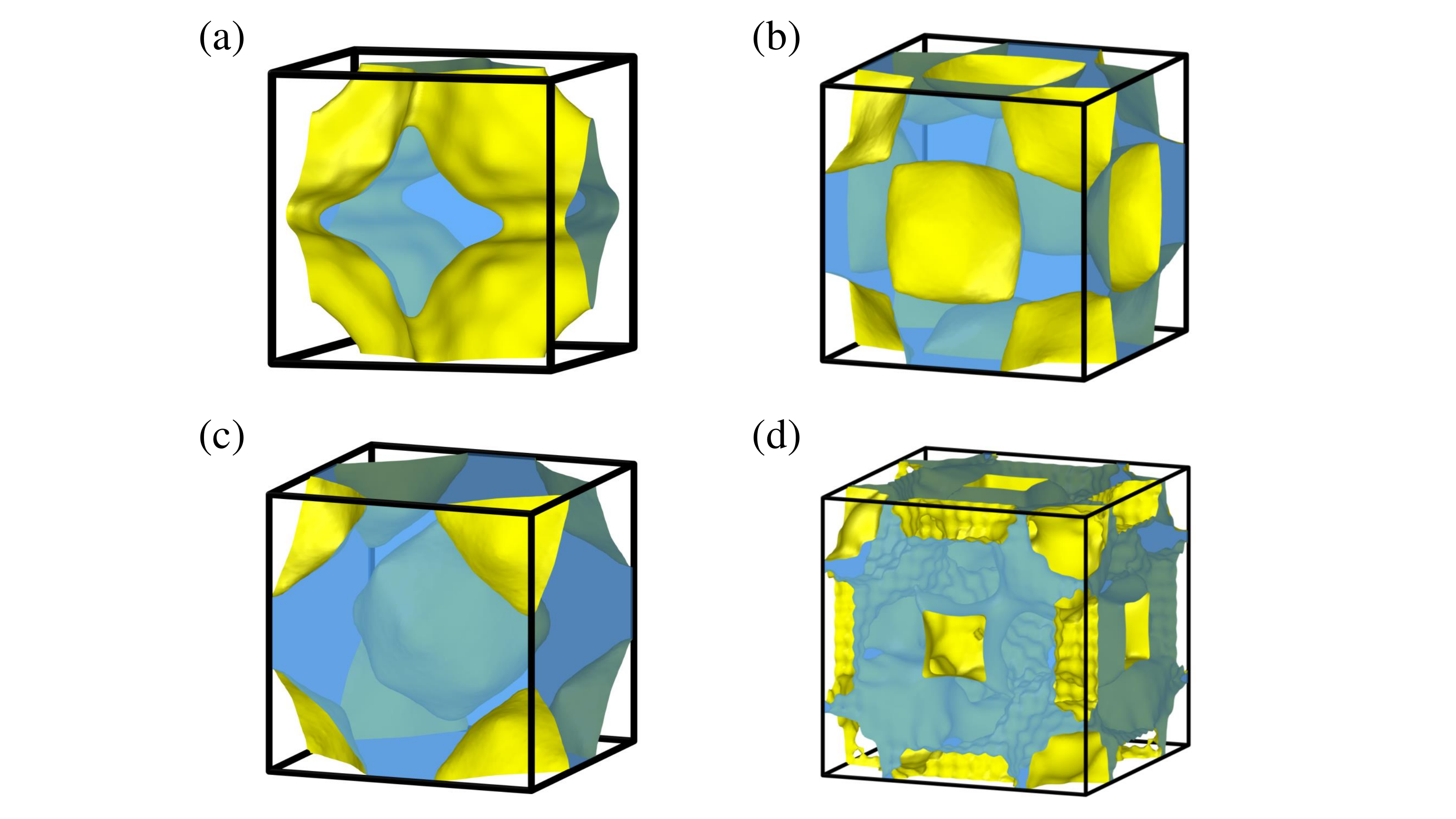}
    	\caption{PANDA visualizations for simple cubic (a), face centered cubic (b), body centered cubic (c), and diamond cubic (d) lattices, respectively. Transparent blue regions display the free volume identified by PANDA, while yellow iso-surfaces indicate the boundary between the volume occupied by atoms and that occupied by free volume. The diamond cubic calculation was performed with $\sim$5 times more voxels compared to body centered cubic, yielding a less smooth contour.}
    	\label{fig:crystal}
\end{figure*}

In this section, we employ PANDA in a proof-of-principle study to characterize the free volume between atoms present in various crystal systems. Figure \ref{fig:crystal} shows a visual depiction of this free volume morphology for simple cubic (SC), face centered cubic (FCC), body centered cubic (BCC), and diamond cubic (DC) crystal structures of Aluminum. All visualization in this work were genereated using the Ovito software package \cite{ovito}. Visual differences clearly exist in the computed free volume topologies, depending on the interstitial sites available within each lattice. For example, as expected the SC lattice indicates a single type of free-volume within the center of the unit cell (blue field in Figure~\ref{fig:crystal}a). The PANDA computed FCC lattice result clearly shows an octahedral interstitial free volume (surrounding the yellow center of the cube face of Figure~\ref{fig:crystal}b) as well as the smaller tetrahedral sites (e.g., in between diagonals connecting atomic volumes, colored yellow). PANDA yields a connected free volume around the central atom in a BCC unit cell (Figure~\ref{fig:crystal}c), as well as the somewhat more complex free volume connecting the DC interstitial sites (Figure~\ref{fig:crystal}d). 

The free volume generated by PANDA can also be compared directly with the true value for each crystal system. By using the atomic packing factor (APF) we can estimate the remaining free volume available in the unit cell. Here we calculate this free volume as V$_{free} = 1 - $ V$_{APF}$, where V$_{free}$ is the available free volume in the unit cell and V$_{APF}$ is the APF for a particular crystal system. PANDA can also be used to calculate V$_{free}$ directly by summing together the volumes of all subgraphs present in the system, as described earlier. Table ST1 shows the PANDA calculated free volume as well as the true free volume calculated from the known APF, for various phases of Aluminum. From Table ST1 one can see the excellent agreement between PANDA and the APF-derived free volumes, indicating that PANDA can reliably capture the unique aspects of free volume topology present in distinct material phases.

\subsection{Autonomous Characterization of Nano-scale Surface Morphology}

We now discuss application of PANDA to understanding the topology of surface features under dynamic conditions. As described in the Computational Details section, PANDA can capture the geometric features of surface structures by generating contour maps of the resulting atom-surface voxel interactions. Conceptually, this is due to the fact that PANDA can be abstracted to examine a variety of different types of interactions, which are ultimately defined by the user. The abstraction of interactions types to graphs provides the user with a nearly limitless number of potential use cases, provided one can construct a set of rules for the graph construction. Here, we showcase this ability by restricting our PANDA analysis to the surface morphology of the aluminum (110) facet during molecular dynamics (MD) simulation of epitaxial growth (containing approximately 250,000 atoms). The MD simulation was performed at 300K for 25 $ns$ with a deposition rate of 1 $\frac{ML}{ns}$. Experimental observations indicate the tendency to form pyramidal structures, where the length of these features are approximately 2-4 times greater than the width \cite{PhysRevLett.91.016102}. All atomic structures used in this section were generated from previous work \cite{doi:10.1021/acs.jpcc.0c05512}. 

\begin{figure*}
        \centering
    	\includegraphics[trim={0 2cm 0 0cm},width=1.0\textwidth]{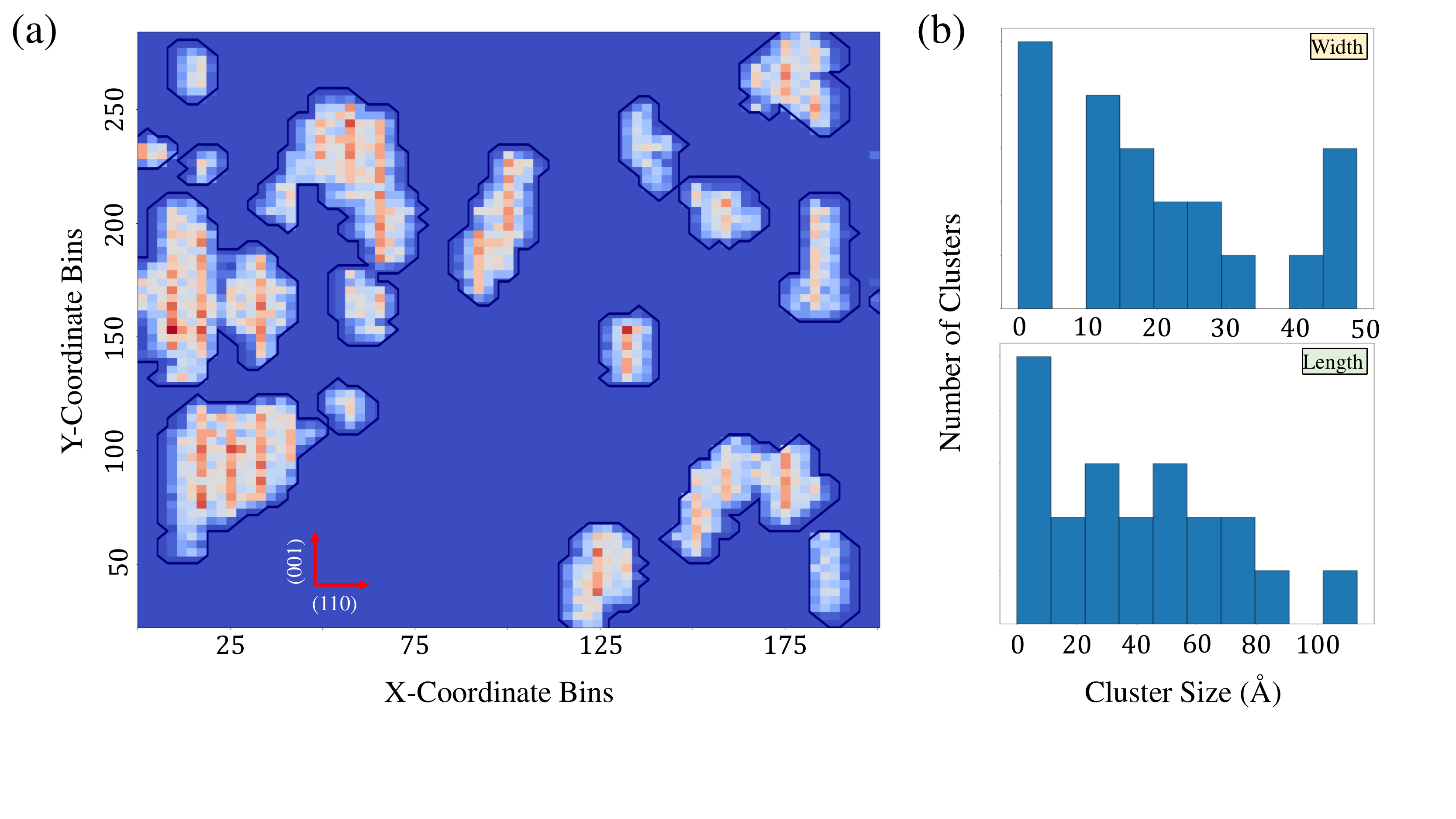}
    	\caption{(a) 2D histogram of the real-space voxel positions. Bin colors represent the average voxel z-coordinate within each bin. Boundary lines represent the contour line that encodes the perimeter of the cluster. The inserted axis, shown in red, indicates crystallographic directions. (b) Histograms of the length and width of the clusters shown in (a). Cluster width and length are discussed in the Computational Details section, which is derived from the contour lines shown in (a).}
    	\label{fig:surface}
\end{figure*}

Figure \ref{fig:surface} (a) shows a contour map of the binned surface axis. Here, the surface features are outlined in black using the graph contouring algorithm described in the Computational Details section. The colors represent the height of the pyramidal structures, with areas of dark blue indicating that no surface features are present above the designated z-axis cutoff. We observe that the PANDA surface map has recovered the pyramidal nature of the surface features, and clearly distinguishes between each unique cluster. The graph contouring algorithm also uniquely identifies where these clusters exist, providing an intuitive and efficient way of calculating each cluster's length and width.

Figure \ref{fig:surface} (b) provide histograms of the length and width of the graph-contoured surface features. We clearly see that the surface cluster length to width aspect ratio of 3:1 falls within the experimentally observed range. We note that the aspect ratio distributions will not necessarily be uniform  due to the limited surface area used during the MD simulations. Regardless, this aspect ratio is only expected to be qualitatively correlated with experimental observations. The ability to employ user-defined rule sets to automatically detect and characterize these types of surface features allows the PANDA algorithm to be utilized for a multitude of materials characterization problems, including nucleation and growth of metal hydrides\cite{Mullen_spin_lattice_2020}, carbon condensation in hot, compressed materials\cite{LCO-2020}, and the study of mass transport in  amorphous solids within grain boundaries of polycrystalline materials\cite{Srolovitz_GB_2009}.

\subsection{Bulk Free-volume Defects}

\subsubsection{Characterization of Nano-scale Free Volume Topology}

\begin{figure*}
        \centering
    	\includegraphics[trim={0 0cm 0 0cm},width=0.8\textwidth]{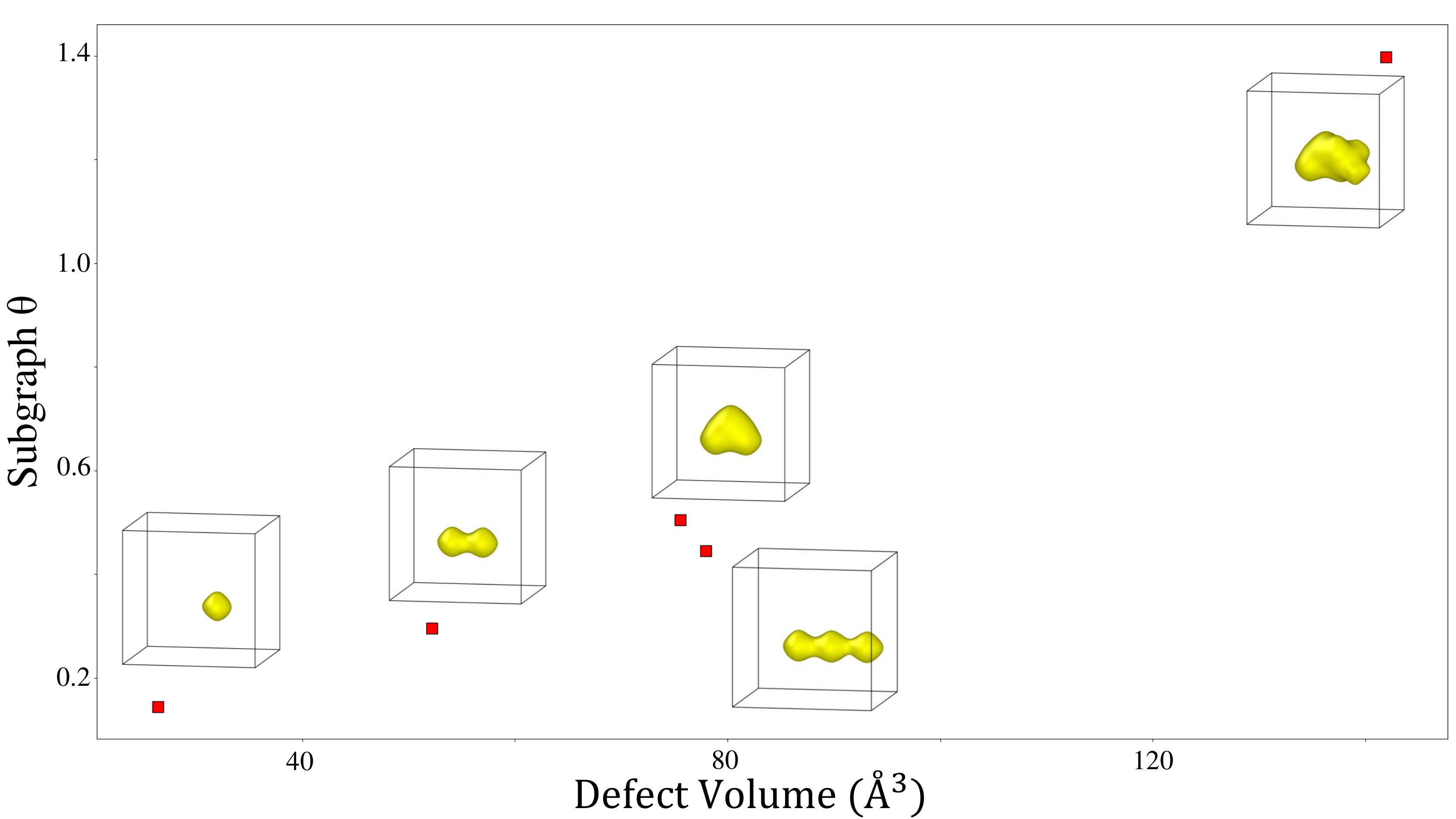}
    	\caption{ Monovacancy, divacancy, two types of trivacancies, and a vacancy cluster in Al. Each free volume's SGOP-V is plotted as a function of the defect's volume. Yellow surface encapsulate the free volume identified by PANDA. }
    	\label{fig:vacs}
\end{figure*}

Here we use a combination of PANDA and a graph-based order parameter (SGOP-V) to map and characterize the subtle differences in the free volume of various types of vacancies and voids. Figure \ref{fig:vacs} shows a visual depiction of this characterization for elemental Aluminum. Figure \ref{fig:vacs} provides the PANDA-generated free volumes for a mono, di, two types of tri-vacancies, and a small vacancy cluster containing five vacancies. One can visually identify the differences between the vacancies from Figure \ref{fig:vacs}, as generated from the inserted PANDA meshes, as well as a strong correlation between the volume of the defects as their corresponding SGOP-V.

From Table ST2, it is clear that the SGOP-V values (generated from PANDA planar graphs) can uniquely characterize different classes of vacancies. The SGOP-V values are also physically meaningful, with a divacancy SGOP-V value being roughly double that of a monovacancy SGOP-V value. One observation to note is that the vacancy cluster containing five vacancies does not yield a SGOP-V value five times that of the monovacancy, which would be the case if all five vacancies were connected in a row, like a chain. However, the vacancy cluster studied here is more densely packed, like a ball, and therefore does not match the shape of five isolated vacancies strung together. This is in contrast to the di- and tri-vacancies, which follow monotonically from the single vacancy SGOP-V value. This physically-informed nature of the SGOP-V provides an intuitive characterization metric where one can capture not only the volume that a particular PANDA graph encompasses, but also its topology.

Figure S1 showcases PANDA's ability to characterize defects in multicomponent systems, namely Al$_{4}$Cu$_{9}$ and Ti$_{11}$Ni$_{9}$Pt$_{4}$. In both systems, a single vacancy of each chemical element is created and a PANDA free volume topology map is generated. Table ST2 provides the SGOP-V values for the various results. We note that values across different material types are not necessarily comparable, as the graph's node degrees are dependent upon the voxel size, which can be different depending on the system. The SGOP-V values from the PANDA map again provide physically intuitive vacancy characterizations. Al and Cu share similar free volumes, though the SGOP-V indicates that Cu does produce a larger free volume overall, commensurate with its larger atomic radius. For the case of Ti$_{11}$Ni$_{9}$Pt$_{4}$, the SGOP-V indicates that the free volume created by Ti, Ni, and Pt vacancies increases respectively, as one would expect due to an increased atomic radius. We note that Ti$_{11}$Ni$_{9}$Pt$_{4}$ does not have an orthogonal set of lattice vectors, and that the PANDA algorithm can be employed to all crystal system.

\begin{figure*}
        \centering
    	\includegraphics[trim={0 0cm 0 0cm},width=0.8\textwidth]{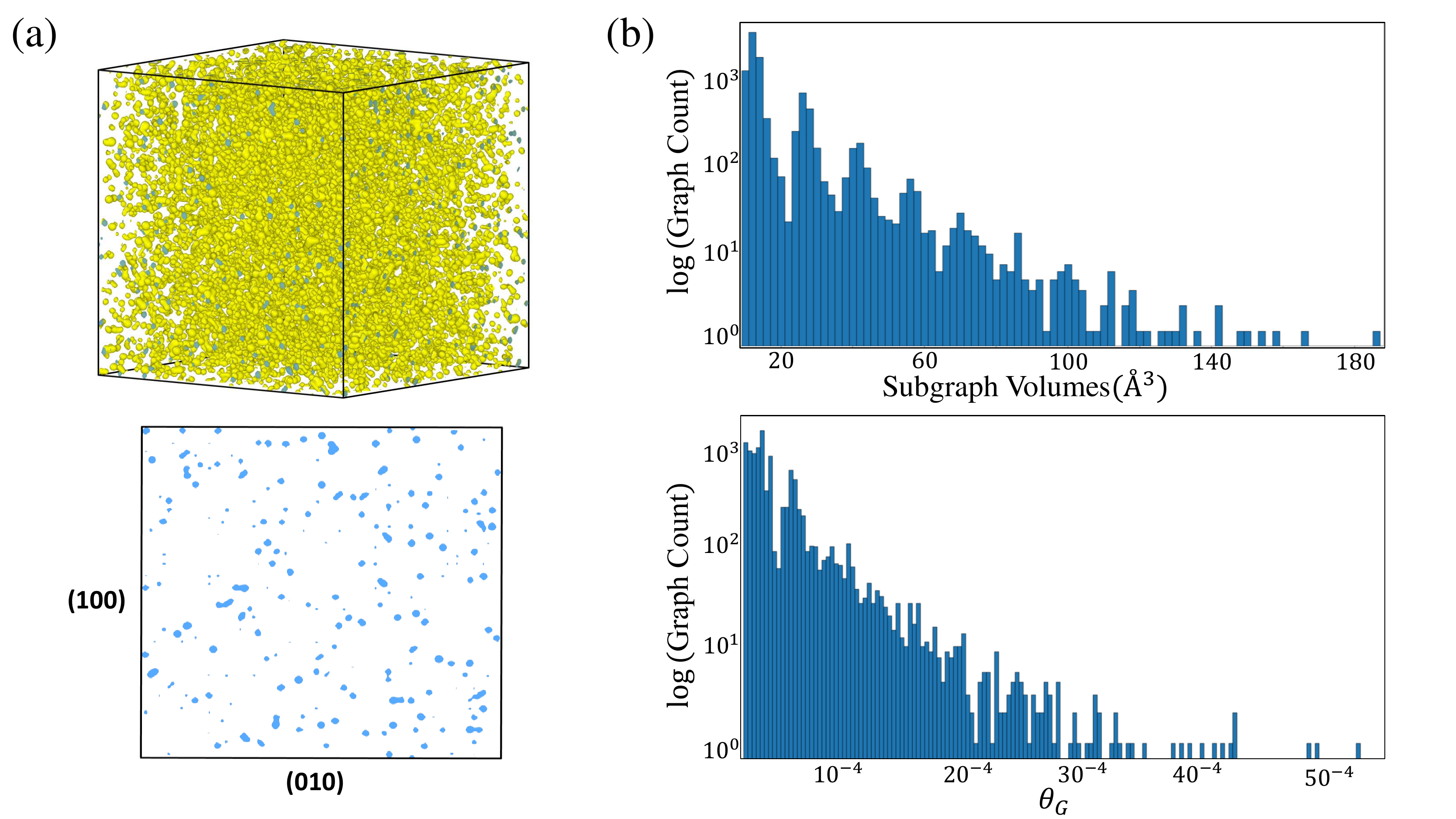}
    	\caption{(a, top) Reconstructed surface mesh overlaying the voxel positions. These surface meshes (yellow) indicate areas of free-volume within the system, with the blue areas indicating periodically replicated free volumes. The shape of the surface mesh is generated from a combination of voxel position and voxel sphere radii. (a, bottom) 1 $\AA$ sliced view of the surface meshes shown above, with the blue regions representing the 2D area of the free-volume defect captured in the sliced view. This view provides the reader with a more intuitive way to visualize the types of defects that are captured in (a, top). (b, top) Histogram indicating the distribution of volumes of each free-volume defect in (a). (b, bottom) Histogram indicating the distribution of SGOP-V values for each free-volume defect's planar graph in (a). }
    	\label{fig:porosity}
\end{figure*}

In order to investigate free volume correlations over larger spatial scales, we employ PANDA to generate a free volume map of elemental Aluminum, spanning over 25 nanometers in each direction. In this case, we generate a porosity of five percent, where atomic sites are randomly removed until the required concentration is satisfied. Figure \ref{fig:porosity} (a, top) provides a visualization of the free volume map for a vacancy concentration of five percent. Figure \ref{fig:porosity} (a, bottom) provides a 2D slice of the overall 3D system. 

Figure \ref{fig:porosity} (b) provides a more quantitative analysis of the resulting defects by observing both the volume and SGOP-V value of each subgraph discovered within the system. Figure \ref{fig:porosity} (b, top) shows a histogram of PANDA planar graph volumes, indicating that the overwhelming majority of defects present in the system are indeed composed of small vacancy clusters. Figure \ref{fig:porosity} (b, bottom) corroborates this by providing a histogram of SGOP-V values on each PANDA-generated subgraph. Here, we observe that the peaks present in both parts of the figure largely overlay, predictably. In addition, each histogram indicates an overall exponential distribution, as expected from randomly chosen vacancy configurations. Some contrast between these two plots occurs where the volume histogram shows a peaked distribution about fixed intervals, corresponding to small vacancy clusters that tend to have similar volumes. These peaks tend to be removed from the SGOP-V histogram, due the fact that different polymorphs of small clusters can have different connectivities/morphologies, e.g., a tri-vacancy can occur as a linear chain or triangle, a tetra-vacancy can occur as a one dimensional chain, two-dimensional, or three-dimensional structure.

\subsubsection{Tracking Vacancy Diffusion Events}

\begin{figure*}
        \centering
    	\includegraphics[trim={0 2cm 0 0cm},width=0.8\textwidth]{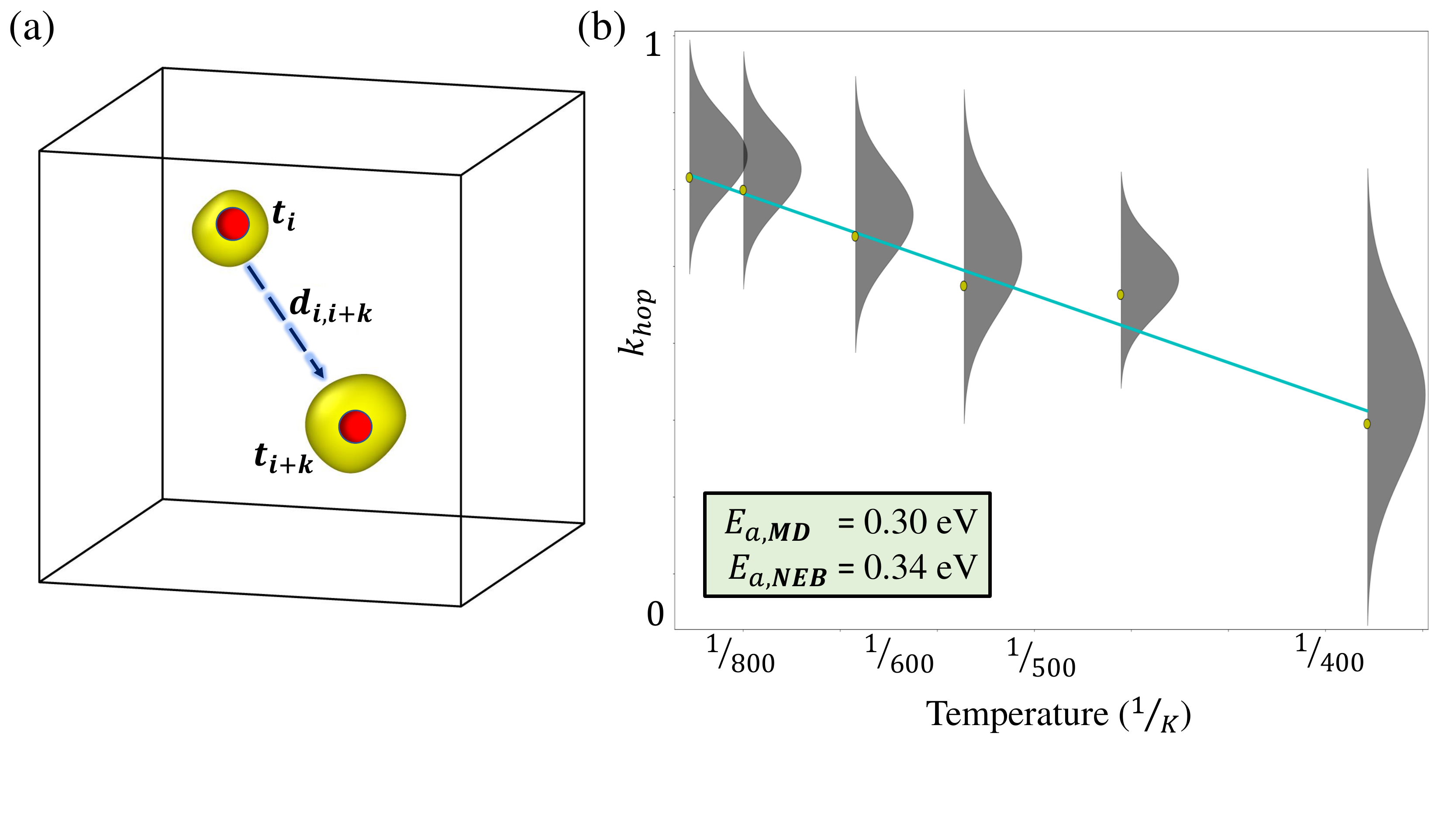}
    	\caption{(a) Visual depiction of determining free-volume event tracking as the free-volume moved from time $t_{i}$ to $t_{i+k}$. Red spheres represent the center of mass of the underlying free-volume graph. Distances, $d_{i,i+k}$, are calculated as the distance between each graph's center of mass. (b) Free-volume displacement rates as a function of temperature. Grey distributions represent the distribution of displacement times over the course of the MD trajectory. Yellow points represent the mean of the displacement rates, which was used to fit the Arrhenius curve (cyan), which was used to calculate the activation energy required for free-volume displacement.}
    	\label{fig:hop}
\end{figure*}

Free volume corresponding to defects present in a material often diffuse through their host under dynamic conditions. Here we use the PANDA planar graph to track diffusion events of a single vacancy in elemental aluminum at various temperatures. The methodology behind this event tracking can be found in the computational details section. To summarize, the planar graph center of mass (COM) can be used to track free volume diffusion by observing when the graph COM changes beyond some threshold. Figure \ref{fig:hop} (a) provides a visual depiction of how the graph COM can be tracked to determine diffusion events. Here, the graph COM (red) at time $t_{i}$, moves by some distance, $d_{i,i+k}$, at time $t_{i+k}$. In principle, multiple graph COM changes can be tracked for various types of free volumes, though in this work we consider only the single vacancy. Here we consider MD simulations of elemental Al using 255 atoms (with a primitive cell optimized by quantum mechanical calculations \cite{doi:10.1021/acs.jpcc.9b03925}) at temperatures ranging from 300K to 900K.

Figure \ref{fig:hop} (b) shows the hop rates calculated from graph COM displacements as a function of temperature. Yellow points represent the average hop rate at a given temperature, while the grey distributions show a Gaussian fit to the hop rates at the specified temperature. The cyan curve was generated from a linear fit to the yellow data points. The activation energy required for vacancy diffusion can then be estimated from an Arrhenius relationship between the hop rate and temperature. The activation energy can also be calculated via a nudged elastic band calculation (NEB) \cite{doi:10.1063/1.1329672} using the Zhou et. al. EAM potential \cite{ZHOU20014005}. Both values are provided in the green insert within Figure \ref{fig:hop} (b). We observe that the PANDA computed activation energy agrees with NEB within 0.04 eV (1 kcal/mol), with differences likely due to entropic effects and the fact that NEB probes the zero temperature potential energy surface, only. NEB calculation also provide no intuitive way to determine uncertainty, whereas our activation energies determined via dynamic phase space sampling inherently provide a means for error estimation. This type of PANDA-computed COM tracking can readily be applied to many temperature dependent phenomena, such as vacancy cluster dissociation, conformational changes of large void systems, as well as the nucleation of free volume defects.

\subsubsection{Spatio-temporal Behavior of Nano-scale Free Volumes}

\begin{figure*}
        \centering
    	\includegraphics[trim={0 0cm 0 0cm},width=1.0\textwidth]{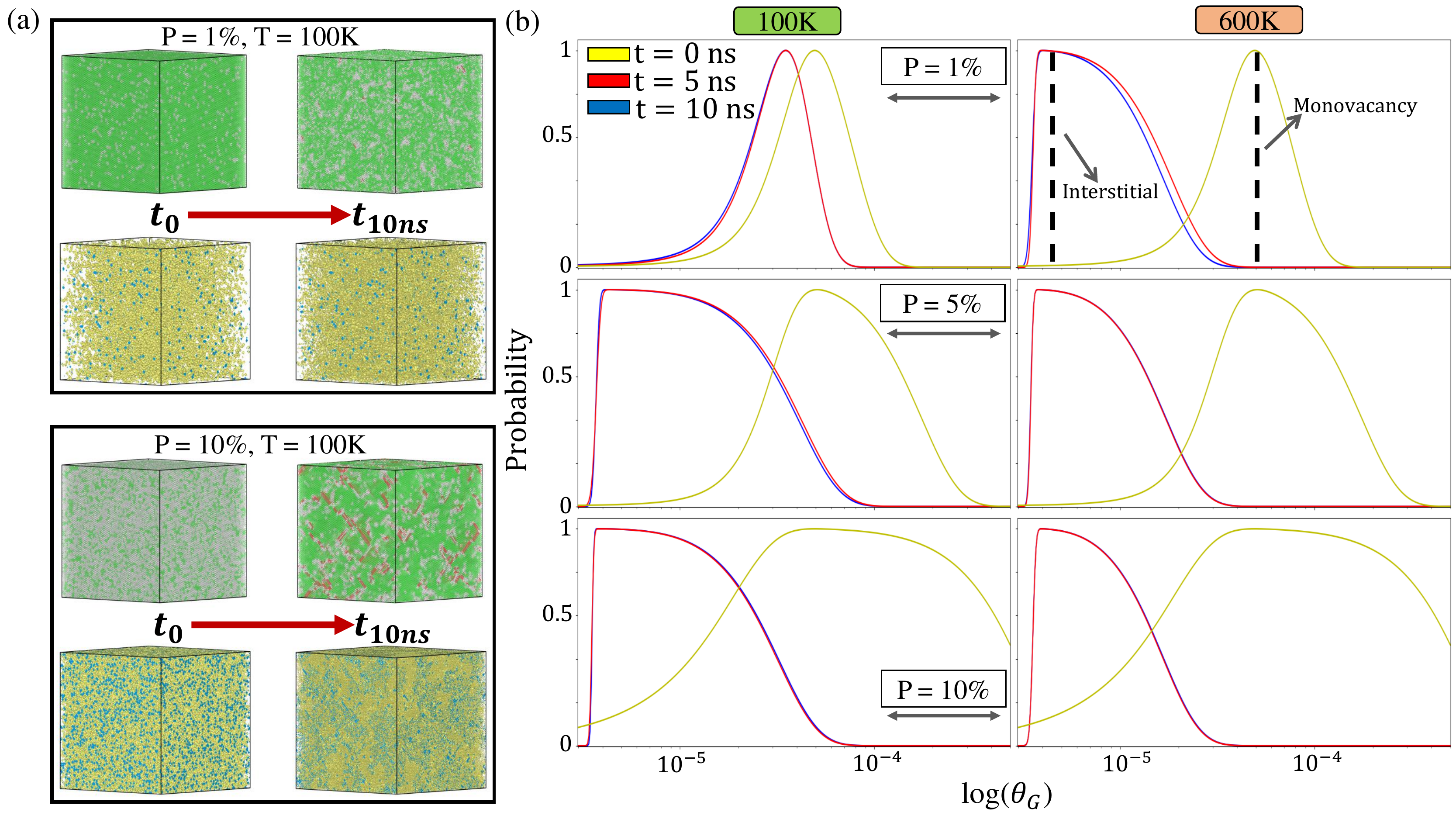}
    	\caption{(a) Qualitative visualization of dynamic free volume evolution using a-CNA (top of each subplot) and PANDA (bottom of each subplot), and (b) quantification of these changes at specific timestamps.  (a,top) For 1\% initial porosity and (a,bottom) 10\% initial porosity, both at T=100K, the a-CNA and PANDA visualizations are shown for $t=0ns$ (right) and $t=10ns$ (left).  (b,top), (b,center), and (b,bottom) show the distributions for 1\%, 5\%, and 10\% initial porosity, respectively, with the left and right columns representing MD runs at T = 100K and 600K, respectively. For reference, in (b,top,right) there are two vertical dashed lines representing the approximate location along the $\theta_{G}$ axis where interstitial and monovacancy environments exist.}
    	\label{fig:por_evol}
\end{figure*}

The spatio-temporal behavior of free volumes is of critical importance, as changes in the underlying free volume topology greatly influence the properties of the material such as stress moduli \cite{KNAUSS1981123}, strength \cite{doi:10.1021/ma060047q}, ductility \cite{Gao_2006}, brittleness \cite{doi:10.1080/01418610208240430}, and other mechanical properties \cite{doi:10.1063/1.5051618}. Here, we explore the time-evolution of several nano-scale free volume morphologies in aluminum via million-atom MD simulations. The finding of these simulations can be found in Fig. \ref{fig:por_evol}. Further details regarding these simulations can be found in the computational details section.

Fig. \ref{fig:por_evol} (a) provides a visualization of the initial and final configurations, with the atomic environments characterized via a-CNA and the free volume with PANDA. Here, one can infer the changes in free volume morphology under dynamic conditions through both the changes in local atomic coordination, through a-CNA (shown at the top of each subplot in (a)), and by observing the differences in the generated PANDA meshes (shown at the bottom of each subplot in (a)). For instance, we can observe that there are only small changes in local atomic coordination in (a,top) for the case of 1\% initial porosity at 100K. We observe a similar trend for the PANDA mesh in (a,top) which highlights that the free volumes present in the initial configuration barely change their size and location within the cell. The opposite trend is observed in (a,bottom), for the case of 10\% initial porosity at 100K, with significant differences in local atomic coordination observed through a-CNA. These changes are also observed within the PANDA mesh, with a clear reduction in the number of larger free volumes and an increase in smaller ones as a function of time. It is important to note that PANDA is not competing against a-CNA, but rather complements it by providing an intuitive explanation for why changes in local atomic coordination occur.

While Fig. \ref{fig:por_evol} (a) aims to qualitatively show the connection between free volume changes and local atomic structure, Fig. \ref{fig:por_evol} (b) quantifies those observations in a physically intuitive manner. From Fig. \ref{fig:por_evol} (b) one can observe the change in free volume topology over time for different initial free volume distributions. Fig. \ref{fig:por_evol} (b,top) shows the change in free volume topology for the case of an initial randomly generated 1\% porosity (where 1\% of the atoms are randomly removed from a perfect FCC system). Fig. \ref{fig:por_evol} (b,top,left) shows a skewnorm fit to the SGOP-V values for three configurations, which represent different snapshots in time during the T = 100K MD simulation. Here, the initial free volume topology, shown in Fig. \ref{fig:por_evol} (b,top,left) as the yellow curve, is centered around a $\theta_{G}$ of $10^{-4}$, with a uniform spread of approximately 1 order of magnitude around the mean. This is consistent with the characterized distributions in Fig. \ref{fig:porosity}, indicating a similar free volume morphology which consists of mostly smaller point defects. After 10ns, one can observe the red and blue curves in Fig. \ref{fig:por_evol} (b,top,left) have shifted to the right, indicating a minor reduction in larger free volumes and an increase in small free volumes such as monovacancies and interstitial environments. 

A different picture emerges for the case of Fig. \ref{fig:por_evol} (b,top,right), which shows the evolution of the 1\% initial free volume morphology at 600K. Here, after 10ns the free volume in the system is composed of small free volumes and a large number of interstitial sites. This implies that above a kinetic threshold the migration of free volumes in the system often lead to dissolution of the free volume, resulting in the creation of abundant interstitial environments. This observation correlates well with experimental observations, which indicate that nano-scale voids in Al tend to dissociate during annealing, leading to a reduction in both the density and size of the free volumes \cite{doi:10.1143/JPSJ.22.319,doi:10.1143/JPSJ.20.965}. While the free volumes studied in this work are roughly 1 order of magnitude smaller than those studied in \cite{doi:10.1143/JPSJ.22.319,doi:10.1143/JPSJ.20.965}, the qualitative trend is quantified through our MD/SGOP-V framework.

This trend continues as one increases the initial porosity of the system, as seen in Fig. \ref{fig:por_evol} (b,center) for the case of 5\% initial porosity. In Fig. \ref{fig:por_evol} (b,center,left), which represents the 100K MD simulations, we can see a much larger spread of the initial free volume distribution, which makes intuitive sense as more atoms have been randomly removed, resulting in larger free volumes. However, after the 10ns MD simulation, the resulting free volume morphology is quite different than that of the 1\% case. Here, the majority of free volume environments in the system are that of interstitial sites, with a minority consisting of small point defects (as indicated by the tail of the fitted distribution at 10ns). This implies that, even at 100K, the system is so energetically unstable that frequently free volume dissolution occurs. Interestingly, this process leads to fewer small free volumes as the system approaches equilibrium than at 1\% porosity, indicating that there is enough kinetic energy in the system upon void dissolution to effectively eliminate the majority of free volumes, leaving only a limited number of smaller voids/point-defects. A similar trend is seen in In Fig. \ref{fig:por_evol} (b,center,right), the 600K case, but with a larger reduction in free volume size and subsequent increase in the likelihood of a free volume defect being an interstitial site. This again makes sense, as the system has enough kinetic energy at 600K to further reduce unstable free volume defects upon the dissolution stage.

For the case of 10\% initial porosity, shown in Fig. \ref{fig:por_evol} (b,bottom), the same trend emerges but again with a more dramatic reduction in free volume after 10ns. For both the 100K and 600K MD simulations, there are effectively no large free volume defects present in the system, and only a limited number of small free volumes such as monovacancies, with the majority off free volume composed of interstitial environments. As ones moves from 5\% porosity and 100K to 10\% posority and 600K, one can observe a consistent reduction in the number of larger free volumes and a subsequent growth in the number of interstitial environments. This is again observed experimentally \cite{doi:10.1143/JPSJ.22.319,doi:10.1143/JPSJ.20.965}, and qualitatively matches our observations. 

The growth of these interstitial sites ultimately leads to lattice mismatches and the nucleation of HCP regions within the material. This can be observed for the case of 10\% posority and 100K in \ref{fig:por_evol} (a,bottom). Here, using a-CNA we observe the growth of HCP sites, shown in red, after 10ns, along with an increase in the number of FCC sites, shown in green, indicating the removal of larger free volume defects (which can be inferred from the distribution of white atoms at $t=0$). This phenomena is also identified using the PANDA mapping in \ref{fig:por_evol} (b,bottom), with a larger increase in the number of small free volumes and a reduction of larger ones. Interestingly, one can qualitatively identify regions of HCP-FCC mismatch by observing areas of concentrated small free volumes, indicating that regions of the material with a high concentration of interstitial-like free volumes are associated with interfacial regions.

\section{Discussion}

Characterizing the morphology of defects in materials is critical to our understanding of the material's underlying properties. Free volume within a material is often correlated with atomic-level transport, as voids and channels can serve as diffusion highways. Mechanical properties such as toughness and fatigue strength are also correlated with free volume, as these sites often serve as stress-points that can eventually lead to failure. Many techniques have been created in the last few decades to explore the topology of free volume defects in materials, such as ML3D, VROI, VOX, and MT. The PANDA algorithm provides a new capability to understand the morphology of these defect environments with a high level of precision by combining the fidelity of voxelization with the characterization power of graph theory. The PANDA algorithm overcomes limitations of these methods by computing the shape and connectivity of a void space over arbitrarily long distances, rather than focusing exclusively on local free volume configurations. We applied PANDA to a variety of distinct materials problems that uniquely employ our method to help facilitate the understanding of point defect characterization, diffusion event tracking, the determination of surface morphology, and the spatio-temporal behavior of free volume topology as a function of both temperature and porosity.

While the PANDA algorithm does require a minimal set of user-adjusted parameters, they are generally chosen in a physically intuitive manner, requiring only a limited understanding of the local geometric structure of the system (such as pair correlation functions). Our free volume to graph representation allows for the physically informed characterization of these defects through the SGOP-V, providing a mathematically robust morphological feature representation. The computational efficiency, algorithmic simplicity, and mathematical robustness ensure that PANDA can accurately explore a materials free volume defect space for any material system. PANDA thus holds promise as a novel computational capability that can have impact on any number of materials problems, including studies of disordered systems, including porous and/or amorphous structures, that are historically difficult to interpret based on experimental data alone. 

The results presented within this work indicate PANDA's potential use in a broad number of potential applications including understanding the nucleation and growth of defects under extreme conditions, dynamic site identification to better understand transport phenomena, and quantifying the intricate spatio-temporal relationship between free volume morphology and local atomic structure. These application domains represent historically challenging spaces for experiments due to the length and time scales often required to reliably quantify them. Therefore, PANDA could serve as a natural bridge between experimental and computational studies, allowing for a physically intuitive and accurate coupling of these two regimes.

\section*{Data Availability}
All data required to reproduce this work can be requested by contacting the corresponding author. 

\section*{Code Availability}

The PANDA software will be made available upon request.

\section*{Acknowledgements}
This work was performed under the auspices of the US Department of Energy by Lawrence Livermore National Laboratory under contract No. DE-AC52-07NA27344.

\section*{Author Contributions}

N. Goldman supervised the research. J. Chapman created the software, designed the methodology, and performed all data analysis. J. Chapman wrote the manuscript with inputs from N. Goldman.

\section*{Competing Interests}
The authors declare no competing financial or non-financial interests.

\printbibliography

\end{document}


\maketitle

\clearpage

\begin{table}
\centering
    \caption{Free volume present in various crystal phases of aluminum. Free volume is calculated as a function of the atomic packing factor.  V$_{free} = 1 - $ V$_{APF}$, while V$_{PANDA}$ is calculated directly from the graph representation.}
\resizebox{6cm}{!}{
	    \begin{adjustbox}{width=\textwidth}
		    \begin{tabular}{c c c}
    			\hline
    			Phase & V$_{free}$ & V$_{PANDA}$  \\
    			\hline
    			FCC   & 0.26 & 0.25 \\
 			BCC   & 0.32 & 0.34 \\
 			SC  & 0.48 & 0.47\\
			DC   & 0.66 & 0.68 \\

    			\hline
		    \end{tabular}
    	\end{adjustbox}
}
\label{tab:table_phases}
\end{table}

\begin{table}
\centering
    \caption{SGOP-V values for a variety of distinct free volume defects in elemental Al, Al$_{4}$Cu$_{9}$, and Ti$_{11}$Ni$_{9}$Pt$_{4}$. A distinct PANDA parameter set was chosen for each system, so values are not expected to be comparable between the systems.   }
\resizebox{6cm}{!}{
	    \begin{adjustbox}{width=\textwidth}
		    \begin{tabular}{c c c}
    			\hline
    			System & Vacancy Type & $\theta_{G}$   \\
			\hline
			Al & & \\
			& Monovacancy & 0.14 \\
			& Divacancy & 0.29 \\
			& Trivacancy & 0.51 \\
			& Vacancy Cluster & 1.38 \\
			Al$_{4}$Cu$_{9}$ & & \\
			& Al-vacancy & 0.004 \\
			& Cu-vacancy & 0.005 \\
			Ti$_{11}$Ni$_{9}$Pt$_{4}$ & & \\
			& Ti-vacancy & 0.002 \\
			& Ni-vacancy & 0.005 \\
			& Pt-vacancy & 0.01 \\
			\hline
		    \end{tabular}
    	\end{adjustbox}
}
\label{tab:vacs}
\end{table}

\begin{figure*}
        \centering
    	\includegraphics[trim={0 0cm 0 0cm},width=1.0\textwidth]{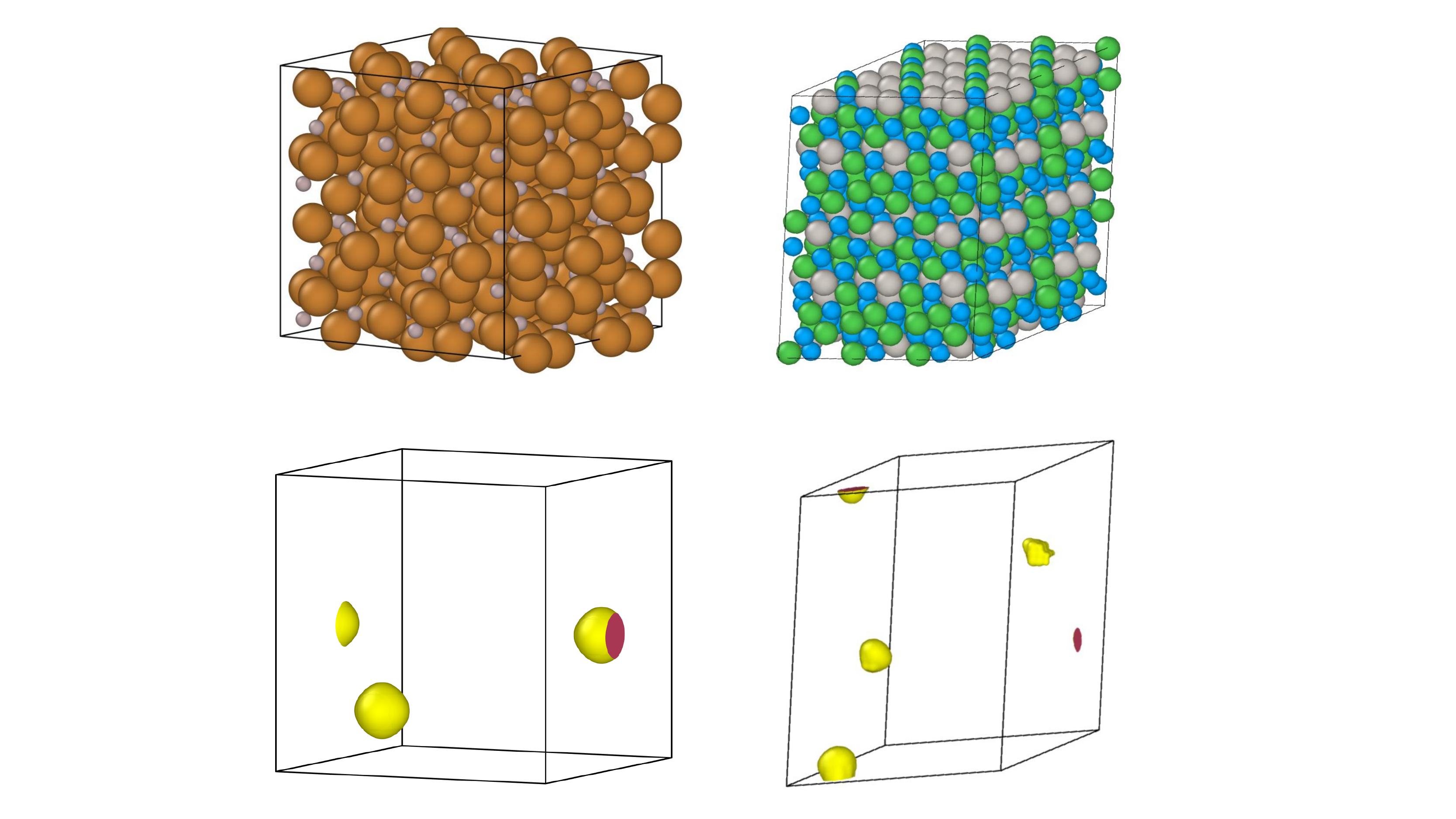}
    	\caption{ (top) Visualization of atomic structure of Al$_{4}$Cu$_{9}$ (left) and Ti$_{11}$Ni$_{9}$Pt$_{4}$ (right). (bottom) PANDA characterized vacancy typologies for vacancies of distinct chemical species. Yellow spheres in the PANDA visualization indicate the identified free volumes, which red shaded areas indicate the periodic boundary of the free volume.}
    	\label{fig:me_vacs}
\end{figure*}

\clearpage

\section*{Acknowledgements}
This work was performed under the auspices of the US Department of Energy by Lawrence Livermore National Laboratory under contract No. DE-AC52-07NA27344.

\printbibliography